\documentclass[12pt,preprint]{aastex}

\newcommand{\bdv}[1]{\mbox{\boldmath$#1$}}

\def\max{{\rm max}}
\def\rel{{\rm rel}}
\def\e{{\rm E}}
\def\bpi{{\bdv\pi}}
\def\bmu{{\bdv\mu}}
\def\bDelta{{\bdv\Delta}}
\begin{document}
\title{Cheap Space-Based Microlens Parallaxes for High-Magnification Events}

\author{Andrew Gould \& Jennifer C.\ Yee}
\affil{Department of Astronomy, Ohio State University,
140 W.\ 18th Ave., Columbus, OH 43210, USA; 
gould,jyee@astronomy.ohio-state.edu}

\begin{abstract}

We show that for high-magnification $(A_\max \ga 100)$ microlensing
events, accurate microlens parallaxes can be obtained from 
three or fewer photometric measurements from a small 
telescope on a satellite
in solar orbit at ${\cal O}$(AU) from Earth.  This is 1--2 orders of
magnitude less observing resources than are required for standard
space-based parallaxes.  Such microlens parallax measurements would
yield accurate mass and distance measurements to the lens for all
cases in which finite-source effects were observed from the ground
over peak.  This would include virtually all high-magnification
events with detected planets and a substantial fraction of those
without.  Hence it would permit accurate estimates of the Galactic
distribution of planets.

\end{abstract}


\section{{Introduction}
\label{sec:intro}}

Microlens parallaxes $\bpi_\e$ can in principle be measured either from the
ground \citep{gould92} or by combining space-based and ground-based
observations \citep{refsdal66}.
The magnitude of $\bpi_\e$ directly yields the ratio of 
the lens-source relative parallax $(\pi_\rel\equiv {\rm AU}[D_L^{-1}-D_S^{-1}])$
to the lens mass $(M)$
\begin{equation}
\pi_\e^2 = {\pi_{\rm rel}\over \kappa M},
\qquad
\kappa\equiv {4 G\over c^2 {\rm AU}} = 8.1 {{\rm mas}\over M_\odot},
\label{eqn:piedef}
\end{equation}
while the direction is that of lens-source relative proper motion, 
$\bmu_{\rm rel}$.  Moreover, if the angular Einstein radius $\theta_\e$
is measured, then the combination of $\theta_\e$ and $\pi_\e$
yields both the mass and the
relative parallax \citep{gould92,gould00}
\begin{equation}
M = {\theta_\e\over \kappa\pi_\e},
\qquad
\pi_\rel = \pi_\e\theta_\e
\label{eqn:masspirel}
\end{equation}

However, both ground-based and space-based microlensing parallaxes
face severe challenges.  To measure parallax from a single location
(e.g., the Earth), the platform must undergo significant deviation
from rectilinear motion: otherwise the platform motion can just be
absorbed into $\bmu_\rel$.  Since, most microlensing events have Einstein
timescales $t_\e$ of weeks, the parallax effects are usually too small
to make useful measurements.  

Space-based measurements work by a substantially different principle.
In essence, one measures the basic lens parameters $(t_0,u_0,t_\e)$
from both the Earth and the satellite.  Here, $t_0$ is the time of closest
approach and $u_0$ is the impact parameter between the source trajectory
and the lens.  One then determines, in effect,
the displacement in the projected Einstein ring between the two 
observatories
\begin{equation}
\bDelta {\bf u} = (\Delta\tau,\Delta u_0),
\quad \Delta\tau\equiv {t_{0,\rm sat} - t_{0,\oplus}\over t_\e},
\quad  \Delta u_0\equiv \pm (u_{0,\rm sat} \mp t_{0,\oplus})
\label{eqn:deltau}
\end{equation}
and then simply divides by the known separation $D_{\rm sat}$ to the
satellite (projected on the plane of the sky)
\begin{equation}
\bpi_\e = {{\rm AU}\over D_{\rm sat}}\bDelta {\bf u}
\label{eqn:piespace}
\end{equation}
The direction of $\bpi_\e$ is thus given relative to the Earth-satellite
vector.  Note that this is conceptually the same as ``terrestrial
parallax'' in which simultaneous observations from different sites
on {\it the Earth} can yield a parallax measurement \citep{hardy95,holz96}.
However, because of the short baseline, this technique can only be 
applied in practice to very rare ``extreme microlensing events''
\citep{gould97,ob07224}.

Equation (\ref{eqn:deltau}) makes clear the principal challenges for
space-based parallaxes.  First, the source must be monitored from
the satellite over many epochs to determine the event parameters that
enter Equation (\ref{eqn:deltau}).  Since satellite time is expensive,
this can be expected to restrict the total number of events measured.

Second, as presented here, the method yields a four-fold degeneracy, which
is the product of two two-fold degeneracies:
the inner ``$\mp$'' in Equation (\ref{eqn:deltau}) depends on whether
the Earth and satellite lie on same or opposite sides of the lens,
and the outer ``$\pm$'' depends on whether the source passes the lens
as seen from the Earth on its left or right (see Figs.\ 1 and 2
of \citealt{gould94}).
Now, the latter degeneracy affects only the direction of $\bpi_\e$, not
its magnitude, and for many applications the direction is of substantially
less interest.  But the former degeneracy does affect the magnitude 
$\pi_\e\equiv |\bpi_\e|$,
and can easily be at the factor $\sim 3$ level.  \citet{gould95} showed
that these degeneracies could in principle be resolved by measuring
the difference in $t_\e$ from the two observatories, but this requires
much higher precision and therefore a several-fold increase in telescope
time, thus gravely exacerbating the challenge that was discussed in
the previous paragraph. 
Now, it is sometimes possible to break these degeneracies 
 by combining Earth-based and satellite-based parallaxes 
\citep{gould99,dong07}, but only for moderately long events.

Here we show that for a subset of microlensing events, those that
peak at high magnification as seen from Earth, excellent measurements
of $|\bpi_\e|$ can be obtained by combining ground-based observations
with a single satellite observation.  These high-magnification events
are of exceptional interest because they are more sensitive to
planetary perturbations \citep{griest98}, 
more likely to yield measurements of $\theta_\e$
(and so, via Equation [\ref{eqn:masspirel}], $M$ and $\pi_\rel$), and
easier to observe.

\section{{Satellite Parallaxes for High-magnification Events}
\label{sec:satpar}}

Consider a single observation of a microlensing event taken
by a satellite at a time when the event is very highly magnified
as seen from Earth.  In particular, suppose that
\begin{equation}
u_\oplus\equiv \sqrt{u_{0,\oplus}^2 + 
\biggl({t-t_{0,\oplus}\over t_\e}\biggr)^2} \ll 1
\label{eqn:udef}
\end{equation}
The satellite observation yields a flux measurement $f_{\rm sat}$.
Using standard techniques (e.g., \citealt{gould10}), one can infer 
the source flux in the satellite band $f_{s,{\rm sat}}$, from the
source flux and color in ground-based bands.  In many cases, particularly
when the satellite measurement is at moderately high magnification and
blending is not severe, it will also be possible to adequately estimate
satellite blended-light flux $f_{b,{\rm sat}}$ by photometric transformation
of the ground-based images.  But even when this is not possible, 
$f_{b,{\rm sat}}$ can be determined from a second satellite measurement
at a much later date.  Then the satellite-based magnification is just
$A_{\rm sat} = (f_{\rm sat} - f_{b,{\rm sat}})/f_{s,{\rm sat}}$, and so
(assuming the lens can be approximated as a point-lens), the
satellite position in the Einstein ring can be calculated from the
inverse of the standard \citet{einstein36} formula
\begin{equation}
u_{\rm sat} = \sqrt{2[(1-A_{\rm sat}^{-2})^{-1/2}-1]} \rightarrow A_{\rm sat}^{-1},
\label{eqn:usatdef}
\end{equation}
where the last limit applies when $A_\max\ga 2$.

\begin{figure}
\plotone{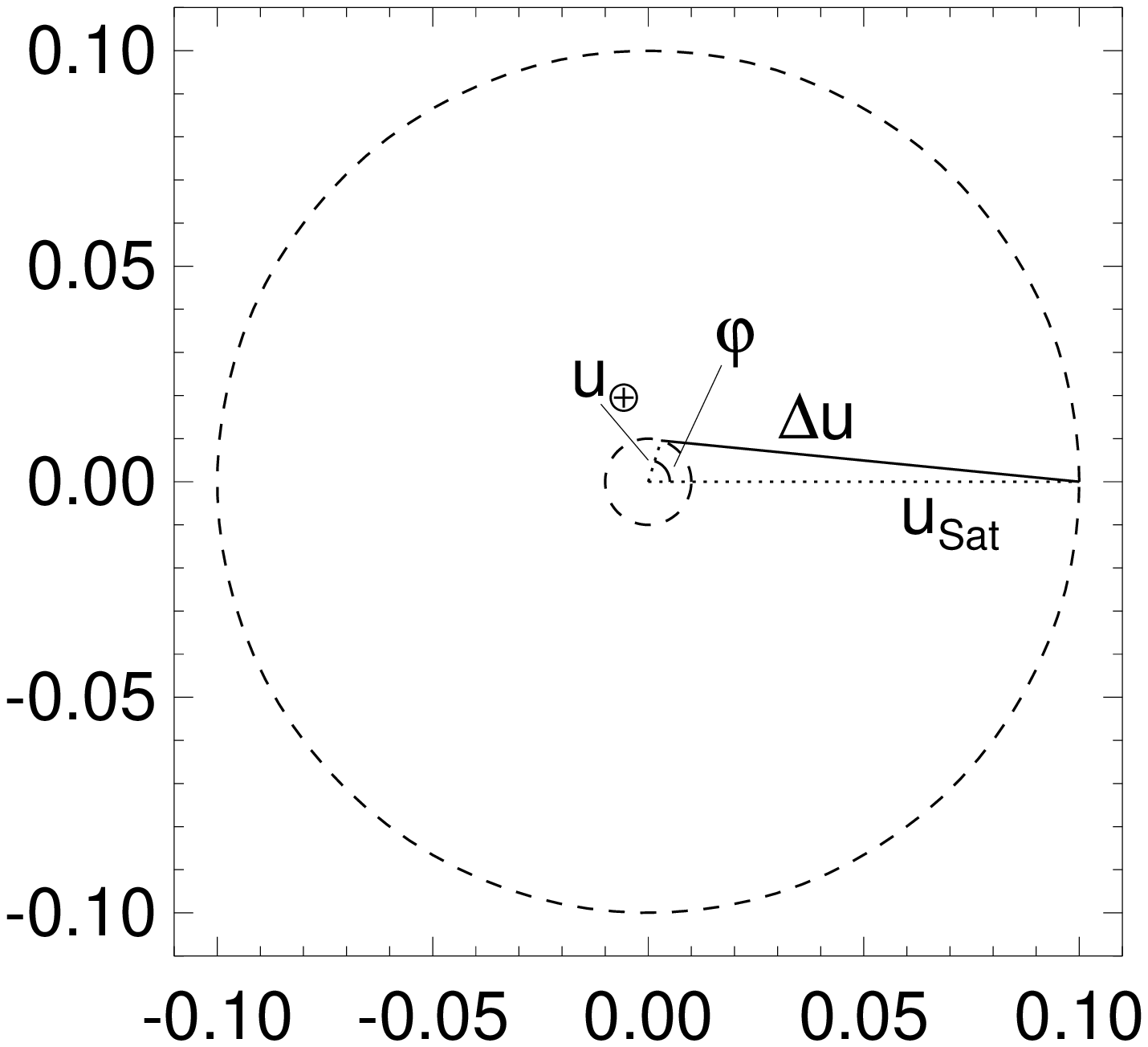}
\caption{\label{fig:ring}
The two circles represent the constraints on the position of
the source relative to the lens (at the origin) as seen from the
Earth, $u_\oplus$, and the satellite, $u_{\rm sat}$. In principle, any
line connecting the inner and outer circles is allowed; one example is
shown. The unknown angle between the lens-source separation vectors as
seen from the satellite and the Earth at the time of the observations
is given by $\phi$. This figure is scaled such that 1.0 equals the
Einstein radius.
}
\end{figure}

By the law of cosines
\begin{equation}
|\bDelta {\bf u}| = \sqrt{u_{\rm sat}^2 + u_\oplus^2 -2 u_{\rm sat}u_\oplus\cos\phi}
\rightarrow u_{\rm sat} - u_\oplus\cos\phi
\label{eqn:udif}
\end{equation}
where $\phi$ is some unknown angle between the lens-source separation vectors
as seen from the Earth and satellite at the time of the observation, and
where the last limit applies for $u_{\rm sat}\gg u_\oplus$.  
Figure~\ref{fig:ring} illustrates this geometry.

Let us initially assume that there are no measurement errors.  
Then by simply adopting
\begin{equation}
\pi_\e = {{\rm AU}\over D_{\rm sat}}u_{\rm sat}
\label{eqn:pienaive}
\end{equation}
one is making a fractional error in the parallax of only
$\delta\pi_\e/\pi_\e = (u_\oplus/u_{\rm sat})\cos\phi$.
The worst errors will occur when the lens is in the Galactic bulge
and has a relatively large mass.  For example, suppose $M=1\,M_\odot$,
and the distances to lens and source are $D_L=7.5\,$kpc and $D_S=8.5\,$kpc;
and let $D_{\rm sat}=1\,$AU.  Then $\pi_\e = 0.044$, so
$|\delta\pi_\e/\pi_\e| < 0.23 (u_\oplus/0.01)$.  Hence, if the satellite
observation is taken when the Earth-based magnification is high,
$A_\oplus\ga 100$ ($u_\oplus\la 0.01$), 
it is possible in principle to make accurate
parallax measurements even though the parallax is much smaller than has
ever been accurately measured from Earth.
Another way to say this is that the systematic error
(due to unknown relative orientation of the Earth and satellite with respect
to the lens geometry) is $|\delta \pi_\e|< ({\rm AU}/D_{\rm sat})u_\oplus$.
This means that for a given satellite separation,
one can simply choose the events with sufficiently high
magnification to achieve the desired precision. Note that this is a 
{\it hard upper limit} on the systematic error, not a 1-$\sigma$ systematic
error.

We now consider the impact of photometric errors. From the limiting form
of Equation (\ref{eqn:usatdef}), one finds that a fractional flux
error measurement leads to the same fractional error in $u_{\rm sat}$,
assuming $u_{\rm sat}\la 0.5$.  Now, for small $u_{\rm sat}$ (where the
systematic errors are important), the source is highly magnified, so
the photometric errors will generally not compete with the systematic
errors.  On the other hand, at moderate $u_{\rm sat}\sim 0.5$, the
systematic errors will be negligible but the photometric errors could
be significant.  In particular, if the source is very faint, then
it is possible that the measurement will be radically compromised,
although even here, one can obtain significant limits just from
the fact that the source was not strongly magnified.  Finally,
if $u_{\rm sat}\ga 1$, then the measurement is likely to be poor
because, from Equation (\ref{eqn:usatdef}), a small error in $A_{\rm sat}$
leads to a large error in $u_{\rm sat}$.
However, for satellite separations $D_{\rm sat}\la 1\,$AU, such large
$\Delta u$ would correspond to a large parallax, which would increase
the probability that the parallax could be measured from Earth-orbital
parallax or possibly terrestrial parallax (e.g., \citealt{ob07224}).

\section{{Application to Planets}
\label{sec:app}}

High-magnification events are an important channel for finding planets
because planets that are anywhere in the system give rise to a
central caustic near the position of the host star.  Hence, if the
event is known to be approaching high magnification (i.e., very small
projected separation between the lens and source) the probability
that the event will probe the central caustic is high, making it
advantageous to apply limited observing resources to the brief
interval of close passage.  Such events are rare, but the specific
rate of planet detection is high.  For example, \citet{gould10b}
derived planet frequencies from a statistically homogeneous
sample of 13 high-magnification events observed over 4 years, which contained 6
planets.  An important feature of these high-magnification events with planets
is that essentially all of them yield measurements of $\theta_\e$.
This is because the source almost always passes close to or over
the central caustic, giving rise to light-curve deviations that
depend on $\rho=\theta_*/\theta_\e$, the source size in units of the
Einstein radius.  Since $\theta_*$ can be determined from the
dereddened color and magnitude of the source \citep{yoo04}, this yields
$\theta_\e= \theta_*/\rho$.

Now, the basic method outlined in Section~\ref{sec:satpar} 
assumed a point lens in order to derive $u_{\rm sat}$ from $A_{\rm sat}$.
See Equation (\ref{eqn:usatdef}).  However, here we are explicitly
consider non-point lenses.  In general, the magnification
pattern for a planet-star lens is very similar to a point lens over
most of the source plane, and so the same approach as given above
will usually work.  Nevertheless, for planetary events, there is some
finite probability (which can be explicitly 
calculated based on the central-caustic
perturbation detected from the ground) that the satellite measurement
will land on (or very near)
the planetary caustic.  To be conservative, one should
therefore take two measurements separated by a short time
(which, again, can be easily calculated based on the ground-based
detection of the central caustic).  In most
cases, these two measurements, combined with the ground-based data,
will be sufficient to virtually rule out that
the first image was ``corrupted'' by a planetary caustic.  Then,
as discussed in Section~\ref{sec:satpar}, it may also be necessary
to take a third image to determine the blending.

If a large fraction of these events also had measured $\pi_\e$,
one would be able to derive the lens mass and lens-source relative
parallax using Equation (\ref{eqn:masspirel}).  This would greatly
increase the value of such high-magnification planet samples.  In particular,
it would allow one to cleanly distinguish between disk and bulge lenses
(and hence planets) and so enable an estimate of
the Galactic distribution of planets.

\acknowledgments

This work was supported by NSF grant AST 1103471.
J.C. Yee is supported by a National Science Foundation
Graduate Research Fellowship under Grant No.~2009068160.

\end{document}